\documentclass[aps,prl,twocolumn,superscriptaddress,showpacs,floatfix]{revtex4-2}
\usepackage{amsmath,amssymb,amsfonts,amsthm}
\usepackage{graphicx}
\usepackage{bm}
\usepackage{bbm}
\usepackage{color}
\usepackage{dcolumn}   
\usepackage{epstopdf}
\usepackage[caption=false]{subfig}
\usepackage{xr}
\usepackage{color}
\usepackage[colorlinks=true,linkcolor=blue,urlcolor=blue,citecolor=blue]{hyperref}

\begin{document}

 \newcommand{\breite}{1.0} 

\newtheorem{prop}{Proposition}
\newtheorem{cor}{Corollary} 

\newcommand{\be}{\begin{equation}}
\newcommand{\ee}{\end{equation}}

\newcommand{\bea}{\begin{eqnarray}}
\newcommand{\eea}{\end{eqnarray}}
\newcommand{\lt}{<}
\newcommand{\gt}{>} 

\newcommand{\Reals}{\mathbb{R}}     
\newcommand{\Com}{\mathbb{C}}       
\newcommand{\Nat}{\mathbb{N}}       

\newcommand{\id}{\mathbboldsymbol{1}}    

\newcommand{\Real}{\mathop{\mathrm{Re}}}
\newcommand{\Imag}{\mathop{\mathrm{Im}}}

\def\O{\mbox{$\mathcal{O}$}}   
\def\F{\mathcal{F}}			
\def\sgn{\text{sgn}}

\newcommand{\deo}{\ensuremath{\Delta_0}}
\newcommand{\dea}{\ensuremath{\Delta}}
\newcommand{\ak}{\ensuremath{a_k}}
\newcommand{\ad}{\ensuremath{a^{\dagger}_{-k}}}
\newcommand{\sx}{\ensuremath{\sigma_x}}
\newcommand{\sz}{\ensuremath{\sigma_z}}
\newcommand{\spl}{\ensuremath{\sigma_{+}}}
\newcommand{\smi}{\ensuremath{\sigma_{-}}}
\newcommand{\alk}{\ensuremath{\alpha_{k}}}
\newcommand{\bk}{\ensuremath{\beta_{k}}}
\newcommand{\ok}{\ensuremath{\omega_{k}}}
\newcommand{\vd}{\ensuremath{V^{\dagger}_1}}
\newcommand{\vi}{\ensuremath{V_1}}
\newcommand{\vo}{\ensuremath{V_o}}
\newcommand{\zc}{\ensuremath{\frac{E_z}{E}}}
\newcommand{\xc}{\ensuremath{\frac{\Delta}{E}}}
\newcommand{\xd}{\ensuremath{X^{\dagger}}}
\newcommand{\aok}{\ensuremath{\frac{\alk}{\ok}}}
\newcommand{\tpw}{\ensuremath{e^{i \ok s }}}
\newcommand{\tpe}{\ensuremath{e^{2iE s }}}
\newcommand{\tmw}{\ensuremath{e^{-i \ok s }}}
\newcommand{\tme}{\ensuremath{e^{-2iE s }}}
\newcommand{\epls}{\ensuremath{e^{F(s)}}}
\newcommand{\emis}{\ensuremath{e^{-F(s)}}}
\newcommand{\epl}{\ensuremath{e^{F(0)}}}
\newcommand{\emi}{\ensuremath{e^{F(0)}}}

\newcommand{\lr}[1]{\left( #1 \right)}
\newcommand{\lrs}[1]{\left( #1 \right)^2}
\newcommand{\lrb}[1]{\left< #1\right>}
\newcommand{\nbt}{\ensuremath{\lr{ \lr{n_k + 1} \tmw + n_k \tpw  }}}

\newcommand{\om}{\ensuremath{\omega}}
\newcommand{\dw}{\ensuremath{\Delta_0}}
\newcommand{\wbp}{\ensuremath{\omega_0}}
\newcommand{\dv}{\ensuremath{\Delta_0}}
\newcommand{\vbp}{\ensuremath{\nu_0}}
\newcommand{\vplus}{\ensuremath{\nu_{+}}}
\newcommand{\vminus}{\ensuremath{\nu_{-}}}
\newcommand{\wplus}{\ensuremath{\omega_{+}}}
\newcommand{\wminus}{\ensuremath{\omega_{-}}}
\newcommand{\uv}[1]{\ensuremath{\mathbf{\hat{#1}}}} 
\newcommand{\abs}[1]{\left| #1 \right|} 
\newcommand{\avg}[1]{\left< #1 \right>} 
\let\underdot=\d 
\renewcommand{\d}[2]{\frac{d #1}{d #2}} 
\newcommand{\dd}[2]{\frac{d^2 #1}{d #2^2}} 
\newcommand{\pd}[2]{\frac{\partial #1}{\partial #2}} 
\newcommand{\pdd}[2]{\frac{\partial^2 #1}{\partial #2^2}} 
\newcommand{\pdc}[3]{\left( \frac{\partial #1}{\partial #2}
 \right)_{#3}} 
\newcommand{\ket}[1]{\left| #1 \right>} 
\newcommand{\bra}[1]{\left< #1 \right|} 
\newcommand{\braket}[2]{\left< #1 \vphantom{#2} \right|
 \left. #2 \vphantom{#1} \right>} 
\newcommand{\matrixel}[3]{\left< #1 \vphantom{#2#3} \right|
 #2 \left| #3 \vphantom{#1#2} \right>} 
\newcommand{\grad}[1]{{\nabla} {#1}} 
\let\divsymb=\div 
\renewcommand{\div}[1]{{\nabla} \cdot \boldsymbol{#1}} 
\newcommand{\curl}[1]{{\nabla} \times \boldsymbol{#1}} 
\newcommand{\laplace}[1]{\nabla^2 \boldsymbol{#1}}
\newcommand{\vs}[1]{\boldsymbol{#1}}
\let\baraccent=\= 

\title{Learning shadows to predict quantum ground state correlations}

\author{Pierre-Gabriel Rozon}
\email{pierre-gabriel.rozon@mcgill.ca}
\affiliation{Department of Physics, McGill University, Montr\'{e}al, Qu\'{e}bec H3A 2T8, Canada}
\author{Kartiek Agarwal}
\email{kagarwal@anl.gov}
\affiliation{Material Sciences Division, Argonne National Laboratory, Lemont, IL 60453, USA}

\date{\today}
\begin{abstract}
We introduce a variational scheme inspired by classical shadow tomography to compute ground state correlations of quantum spin Hamiltonians. Shadow tomography allows for efficient reconstruction of expectation values of arbitrary observables from a bag of repeated, randomized measurements, called snapshots, on copies of the state $\rho$. The prescription allows one to infer expectation values of $M$ $k-$local observables to accuracy $\epsilon$ using just $N \sim 3^k \text{log}M /\epsilon^2$ snapshots when measurements are performed in locally random bases. Turning this around, a bag of snapshots can be considered an efficient representation of the state $\rho$, particularly for estimating low-weight observables, such as terms in a local Hamiltonian needed to estimate the energy. Inspired by this, we consider a variational scheme wherein a bag of $N$ parametrized snapshots is used to represent the putative ground state of a desired local spin Hamiltonian and optimized to lower the energy with respect to it. Additional constraints in the form of positivity of reduced density matrices, motivated by work in quantum chemistry, are employed to ensure compatibility of the predicted correlations with the underlying Hilbert space. 
Despite the limited set of imposed constraints, we show that the optimized ensemble of snapshots yields fairly accurate estimates of the ground state energy and other correlations, including those that do not appear in the set of constraints used in optimization.
\end{abstract}
\maketitle

\textbf{\textit{Introduction.---}}The efficient computation of many-body ground (and excited) states of large quantum systems represents a formidable barrier to scientific progress across a diverse set of disciplines. It has long been appreciated that the exponential growth of the Hilbert space with system size is potentially misleading---for instance, ground states of Hamiltonians usually have lower entanglement and reside in a vanishingly small subspace of the full Hilbert space \cite{HASTINGS2007, BRANDAO2015, ANSHU2022}; and in most cases, physically relevant properties can be inferred from a vanishingly small class of low-weight observables among all possible observables \cite{HUANG2020, VERSTICHEL2012}. 
The former observation has motivated tensor-network approaches, which provide efficient representations of quantum states in low-dimensional systems by exploiting their entanglement structure \cite{VERSTRAETE2006, SCHOLLWOCK2011}.

The latter has inspired reduced density matrix approaches (RDMAs), which abandon the goal of reconstructing all observables and instead variationally optimize low-order correlators—such as the two-particle reduced density matrix—while enforcing representability constraints imposed by the underlying Hilbert space \cite{MAZZIOTTI2002, MAZZIOTTI2005, 
HAMMOND2005,
MAZZIOTTI2005_spin,
MAZZIOTTI2006, ALEXANDER2006,
MAZZIOTTI2007, 
MAZZIOTTI2010, MAZZIOTTI2014,  MAZZIOTTI2020, MAZZIOTTI2024}.   

In this work, we describe a novel variational approach that attempts to improve upon limitations encountered by the above two approaches. Underlying this is the novel approach to quantum state tomography called classical shadows---specifically, it prescribes a randomized measurement scheme on $N$ copies of a quantum state $\rho$ and details how certain correlators can be efficiently retrieved from such measurement data, also called snapshots \cite{HUANG2020}. When all spins in the $N$ copies are measured in Haar-randomized local spin bases, the protocol asserts that $M$ $k$-local operators (that act non-trivially on $k$ spins) can be retrieved with additive error $\epsilon$ using $N \sim 3^k \log M /\epsilon^2 $ snapshots \cite{HUANG2020}. This implies that if one is primarily interested in estimating a polynomial in system size $L$ number of low-weight correlators to within some fixed error---enough to accurately estimate, for instance, the ground state energy density of a gapped system---a logarithmic in $L$ number of snapshots $N$ can be used to efficiently represent the state of the system \cite{HUANG2020, HUANG2021}. Instead of obtaining these snapshots from experimental data, here we parameterize them to represent correlations in an arbitrary state, and optimize the parameters such that the represented state minimizes energy 
with respect to the spin Hamiltonian whose ground state we wish to find.  

\begin{figure}[ht]
    \centering
    \includegraphics[width=0.4\textwidth]{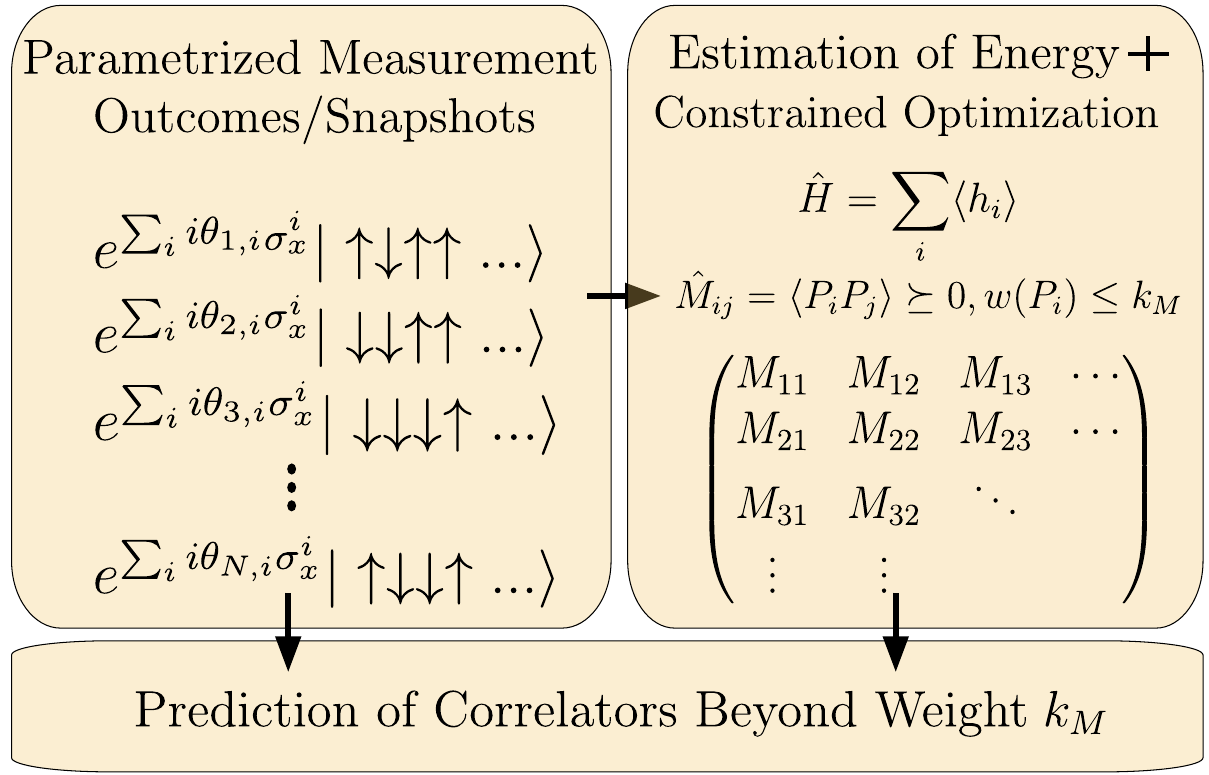}  
    \caption{Variational Scheme: A bag of parameterized snapshots are used to indirectly represent the state of the system (see main text). 
    This ensemble of snapshots can be used to infer terms in the Hamiltonian 
    and pairwise Pauli correlators $\operatorname{Tr}(\rho P_{\vec{b}}P_{\vec{a}})$ of weight $ w \le 2 k_M$ via the shadow inversion procedure. The parameters $\theta_{j,l}$ describing the snapshots are optimized to reduce the estimated energy $\hat{H}$ while enforcing the positivity of $\hat{M}$. Upon optimization, the bag of snapshots can also be used to estimate correlators outside of $H, M$.}    
\label{fig:Scheme}
\end{figure}

Our proposed ansatz takes inspiration from the RDMA \cite{MAZZIOTTI2002, MAZZIOTTI2005, 
HAMMOND2005,
MAZZIOTTI2005_spin,
MAZZIOTTI2006, ALEXANDER2006,
MAZZIOTTI2007, 
MAZZIOTTI2010, MAZZIOTTI2014,  MAZZIOTTI2020, MAZZIOTTI2024},
but enriches their predictive capacity; see Fig.~\ref{fig:Scheme}. In the latter, one optimizes the energy while imposing constraints on low weight correlators built out of operators $O_a \in \{O_1,...\}$ with maximum weight $k_M$, by enforcing the positivity of the correlation matrix $M_{b,a} = \operatorname{Tr}(O_b^{\dagger}\rho O_a)$ comprised of all possible pairs $(b,a)$. While we employ similar constraints as these approaches, the optimized bag of snapshots can be further used to predict higher weight correlators that do not appear in these constraints; numerical evidence shows indeed that the accuracy of predicting such correlators improves with increasing $N$. In this letter, we detail our variational ansatz, discuss its scaling properties, and present data that illustrates its efficacy in predicting correlators of different weights in the ground state of one-dimensional spin models. We conclude by listing promising directions for future refinements of this ansatz and further explorations.    

\textbf{\textit{Variational Scheme.---}} 
The $l^{\text{th}}$ (of $N$) snapshot can be represented as
\begin{align}
\ket{b_l} (\vec{\theta}_l) = \bigotimes_{j = 1}^L e^{i\frac{1}{2}\theta_{j,l}\sigma^x_j}\ket{0} \; \;,  \; \; \mathcal{U}_l = \bigotimes_{j = 1}^L U_{j,l}, 
\end{align}
where $\ket{b_l}$ represents a possible measurement outcome obtained after the putative application of a random unitary $\mathcal{U}_l$ on a copy of the ground state $\rho$ of a Hamiltonian which we wish to study. Here $\vec{\sigma}_j$ are Pauli operators acting on the $j^{\text{th}}$ qubit, $\theta_{j,l} \in [0, \pi]$ parameterize the snapshots, and $U_{j,l}$ are \emph{fixed} single-qubit unitaries drawn from the Haar ensemble. While measurement outcomes in usual shadow tomography are computational basis states $\ket{1001...}$, we allow for smooth variation between the antipodal qubit configurations $\ket{0}$ and $\ket{1}$ to obtain smooth derivatives amenable to optimization procedures. 

In this parametrization each snapshot can be inverted per shadow tomography prescription \cite{HUANG2020} to yield a density matrix estimate, 
\begin{align}
    \hat{\rho}_l = \frac{1}{2^L} \bigotimes_{j = 1}^L \left(I + 3 \vec{n}_{j,l} (\theta_{j,l} ) \cdot \vec{\sigma}_j \right), 
    \label{eq:snapinv}
\end{align}
of the true state $\rho$. Here $I$ is the $2\times2$ identity matrix on one qubit and $\vec{n}_{j,l}$ are given by the local measurement axis determined by $U_{j,l}$ and $\theta_{j,l}$; see SM~\cite{footnote1}. 

The form of $\hat{\rho}_l$, Eq.~(\ref{eq:snapinv}), is the standard result for shadow tomography with measurements in Haar-random bases, and is justified in this setting as long as the unitaries $U_{j,l}$ and angles $\theta_{j,l}$ modulo $\pi$ are statistically independent; see SM~\cite{footnote1} for details.

In the usual setting where an experimentalist obtains snapshots by measuring copies of the unknown state $\rho$, the average $\hat{\rho} = \frac{1}{N} \sum_{l=1}^N \hat{\rho}_l$ is guaranteed to converge to $\rho$. In this variational approach, we are not provided $\rho$ to perform measurements on---instead we optimize $\vec{\theta}_l$ to lower the energy $\text{Tr}\left[ \hat{\rho} H \right]$ of the predicted ground state $\hat{\rho}$ with respect to $H$. 
The bag of snapshots are thus treated as a variational representation of the true state $\rho$. 

It is also worth noting that unlike usual density matrices, $\hat{\rho}_l$ and their average $\hat{\rho}$ are not positive semi-definite (PSD) matrices. Thus, expectation values of observables computed with respect to $\hat{\rho}$ do not necessarily satisfy constraints imposed upon them by the underlying Hilbert space and must be explicitly constrained. (This is counter-intuitively, somewhat fortunate---it would be much harder to obtain a pure state $\rho = \ket{\psi} \bra{\psi}$ by summing over physical mixed density matrices!) 

To this end, we take inspiration from RDMAs \cite{MAZZIOTTI2002, MAZZIOTTI2005, 
HAMMOND2005,
MAZZIOTTI2005_spin,
MAZZIOTTI2006, ALEXANDER2006,
MAZZIOTTI2007, 
MAZZIOTTI2010, MAZZIOTTI2014,  MAZZIOTTI2020, MAZZIOTTI2024} and we enforce positive semi-definiteness of a matrix of correlators 
\begin{align}
\hat{M}_{\vec{b},\vec{a}} = \text{Tr} \left[ \hat{\rho} P^\dagger_{\vec{b}} P_{\vec{a}} \right], 
\end{align} 

where $P_{\vec{a}} = \bigotimes_{j=1}^L \sigma^{a_j} $ with $\vec{a}$ a vector whose elements (either $I,X,Y,Z$) specify the Pauli operator at site $j$ appearing in the tensor product. In what follows, we construct the correlation matrix using all pairwise correlators between Pauli operators of maximum weight $k_M = 2$ and maximum range $r_M \in \{2,3,4\}$ depending on the model and system size $L$. This truncation is analogous in spirit to the 2-RDM constraints used in quantum chemistry, where enforcing positivity conditions on reduced correlation structures is known to yield accurate energies and local observables for Hamiltonians containing at most two-body interactions \cite{MAZZIOTTI2002, MAZZIOTTI2005, 
HAMMOND2005,
MAZZIOTTI2005_spin,
MAZZIOTTI2006, ALEXANDER2006,
MAZZIOTTI2007, 
MAZZIOTTI2010, MAZZIOTTI2014,  MAZZIOTTI2020, MAZZIOTTI2024}.  
The use of such constraints is additionally motivated by the fact that the variance of the classical shadows estimator scales exponentially in the weight of the operator. 

We note that unlike RDMAs where the elements of $M_{\vec{b},\vec{a}}$ are themselves parameters which are optimized over, we estimate $\hat{M}_{\vec{b},\vec{a}}$ from a bag of snapshots which are parameterized by $\vec{\theta}_{l}$. This  allows us, after optimization of $\vec{\theta}_l$, to estimate correlators that are not part of $H$ or $M_{\vec{b},\vec{a}}$. 

The cost function incorporates both the energy term and the enforcement of constraints and is given by 
\begin{align}\label{eq:costfunction}
    f(\vec{\theta}_l, t) = g \hat{H} (\vec{\theta}_1, ..., \vec{\theta}_N) - \mu(t) \text{Tr} \log \left[ \hat{M} (\vec{\theta}_1, ..., \vec{\theta}_N)\right]. 
\end{align}
The logarithmic barrier
$
\mu(t)\sum_i \log\!\left(\frac{1}{\lambda_i}\right)$
diverges as any eigenvalue \(\lambda_i\) of \(\hat{M}\) approaches zero, enforcing positivity.
At training step $t = 0$, the initial random distribution of angles $\theta_{j,l}$ generally produces a correlation matrix $\hat{M}$ containing negative eigenvalues. Since the logarithmic barrier term becomes ill-defined outside the PSD region, an initial feasibility pre-optimization is first performed to shift all eigenvalues above zero (see SM \cite{footnote1}) without the energy term.  
Once the matrix $\hat{M}$ becomes PSD, we set $g = 1/L$, and the barrier coefficient is progressively reduced according to $\mu(t) = \frac{A_{\text{final}}^{t/T}}{\text{dim}(M)}$ with $T = 300$ and $A_{\text{final}} = 5\times10^{-4}$. The optimization is then carried out using gradient descent with adaptive backtracking and feasibility-preserving updates, ensuring that the matrix remains within the PSD region throughout the optimization procedure (see SM \cite{footnote1} for details).

For simplicity, we focus on translationally invariant systems but this invariance is not explicitly imposed. We anticipate that the initial state, being close to the fully mixed state, exhibits approximate translational symmetry. For the translationally invariant models studied, we find the procedure converges to solutions that respect this symmetry.

\begin{figure}[htbp]
    \centering
    \includegraphics[width=0.48\textwidth]{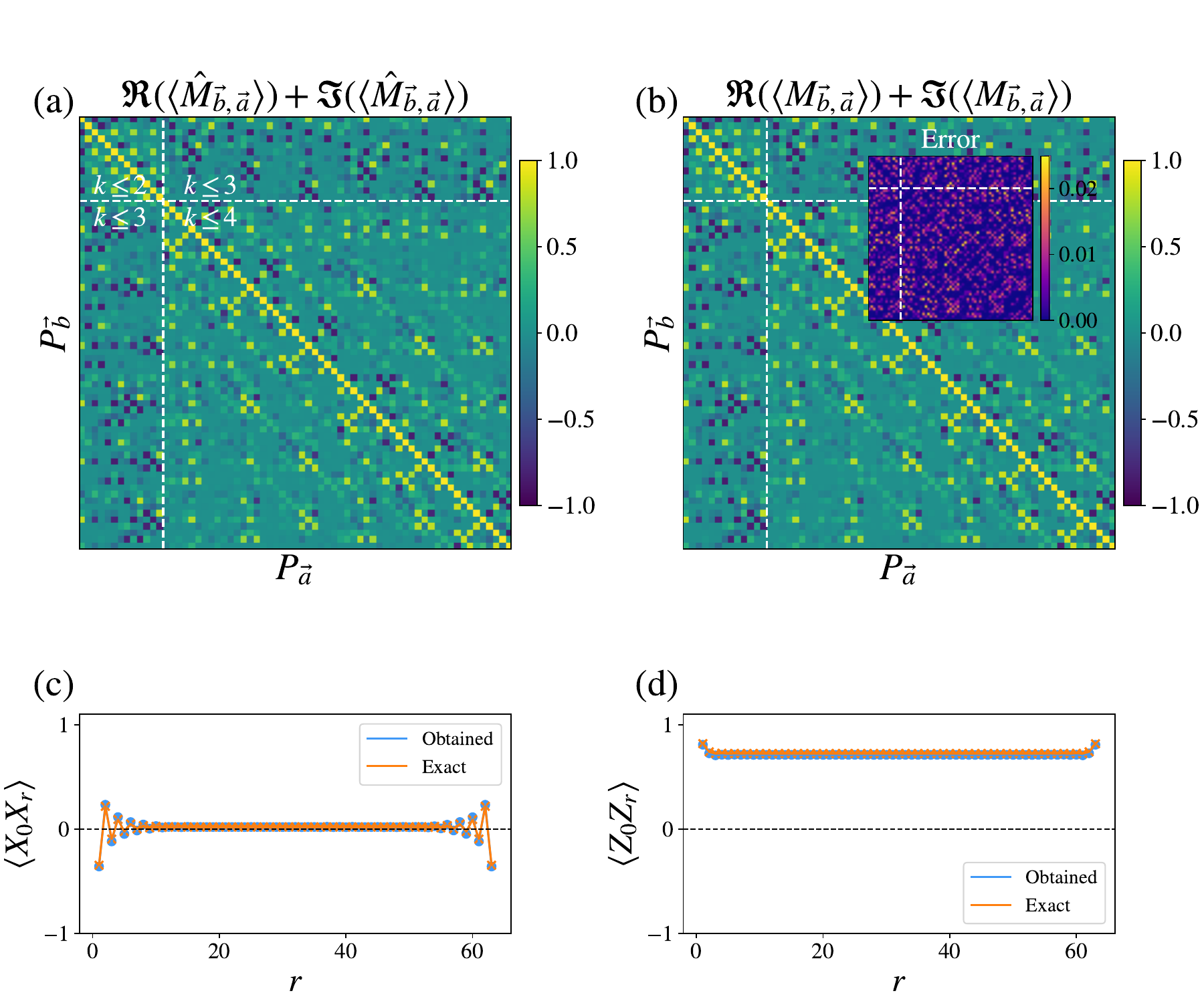}  
    \caption{The sum of real and imaginary parts of the correlation matrix $M_{\vec{b},\vec{a}} = \avg{P_{\vec{b}} P_{\vec{a}}}$ (whose entries are either purely real or purely imaginary) for all Pauli operators $P_{\vec{b}}, P_{\vec{a}}$ of weight $k \le 2$ as obtained (a) from shadow learning for $N = 16000$ and (b) DMRG, for the system governed by Hamitlonian $H_0$}. Figure (a) and (b) display the correlators for a sub-region
    of 4 contiguous qubits, embedded in a system of size $L = 20$. The inset in (b) shows the error. Dashed lines demarcate regions with different maximum weight of the product $P_{\vec{b}} P_{\vec{a}}$. (c) and (d) show the $\avg{Z_0 Z_{r}}$ and $\avg{X_0 X_{r}}$ correlations. `Exact' results were obtained using DMRG.
\label{fig:M}
\end{figure}

\begin{figure*}[ht]
    \centering
    \includegraphics[width=0.9\textwidth,trim=7cm 0cm 6cm 0cm]{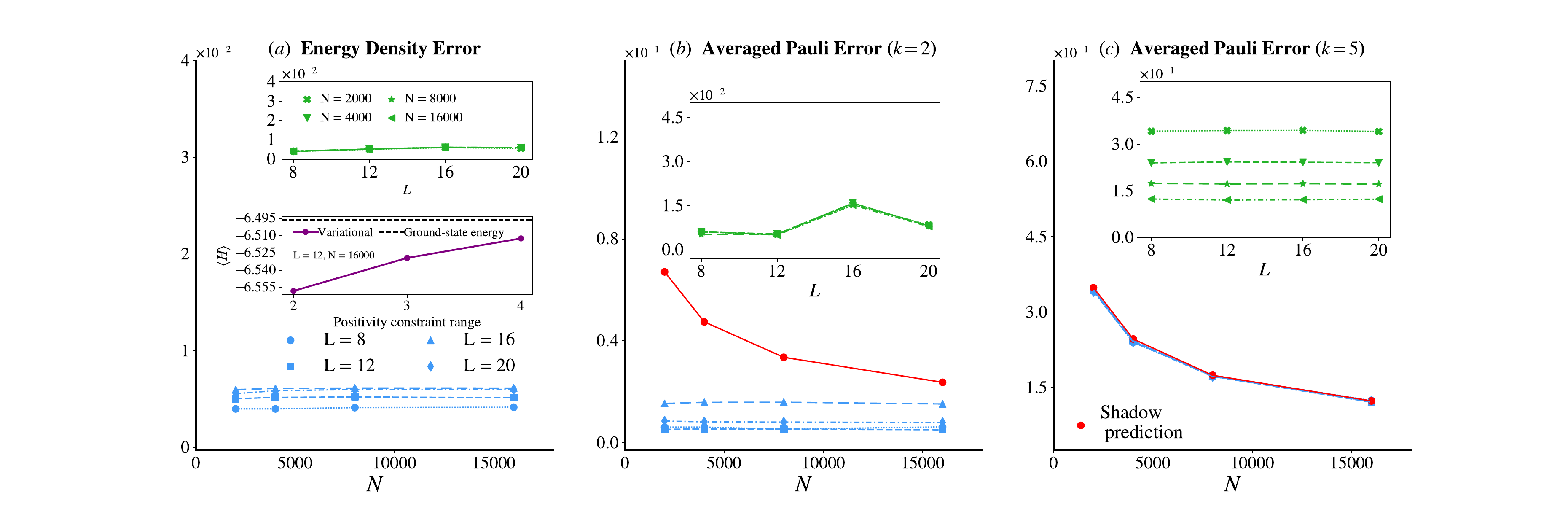}
    \caption{Scaling of error in the (a) energy density, (b) and (c) contiguous Pauli operators of weight $k = 2$ and $k = 5$, respectively, with snapshot number $N$, for the Hamiltonian $H_1$. Errors are computed by comparing the obtained results from snapshot learning and exact sparse matrix diagonalization, as $\epsilon = \sqrt{\left(\sum_{\text{contiguous }P, w(P)  = k} (\hat{P} - \langle P\rangle)^2\right)/3^k}$. Results are presented for positivity constraints with range $r_M = 2$. Insets in green show the same data against system size $L$. The second inset in $(a)$  shows the obtained variational energy compared the exact ground state energy---it is always below the true ground state energy and approaches it as the range of operators in the positivity constraints, $r_M$, is increased. Red lines indicate scaling estimate from shadow tomography for errors in operators of the corresponding weight.
    }
    \label{fig:Scaling}
\end{figure*}

\textbf{\textit{Protocol features.---}}
This protocol differs from usual variational Ansatzes in important ways. First, it is not tied to a specific tensor-network structure, Gaussian representation, or neural-network architecture, and instead relies on a statistical parameterization of the state in terms of measurement outcomes. This implies greater flexibility in representing states with different entanglement and correlation structures, and in higher dimensions. Next, the variational protocol tends to yield states with energies lower than the true ground state energy---this is because the optimization landscape is in fact bounded by the set of constraints used; increasing the set of these constraints results in tighter bounds on the ground state energy; see Fig.~\ref{fig:Scaling}. Thus, the approach provides a (tight, as we find) lower bound on the true ground state energy as opposed to an upper bound. Concomitantly, the choice of constraints inform the biases inherent in the Ansatz, and allows for systematic tradeoffs between computational cost and reconstruction accuracy. Finally, a more subtle aspect of the present variational scheme is that the variance of all Pauli observables of weight $k$ is bounded by $3^{k}/N$~\cite{HUANG2020, agarwal2026learning}.
This stands in contrast to usual Monte-Carlo sampling in wavefunction based Ansatzes, where there are no guarantees on the boundedness of the variance of local estimators. 
\textbf{\textit{Numerical Findings.---}} 
We implement the scheme outlined above to determine ground state correlations for a variety of one-dimensional spin models with periodic boundary conditions. In Fig.~\ref{fig:M}, we present results for an $L = 64$ site system governed by the Hamiltonian $H_0 = \sum_j 0.3\sigma^x_j \sigma^x_{j+1} + 0.4\sigma^z_j + 0.2\sigma_j^x 
$, which is gapped and posseses no spin symmetries. In Fig. \ref{fig:Scaling} we present results for scaling of the error in computing the energy density, arbitrary weight $2$, and higher correlators, with respect to system size and snapshot count,  for the approximately gapless (based on DMRG computation) Hamiltonian $H_1 = \sum_j0.25\, Z_j Z_{j+1}
+ 0.3\, X_j X_{j+1}
+ 0.3\, Y_j Y_{j+1}
+ 0.25\, Z_j
+ 0.3\, X_j$. 

In Figs.~\ref{fig:M} (a) and (b), we show the correlation matrix $\hat{M}_{\vec{b},\vec{a}}$ obtained from learned shadows, and compare it to the result $M_{\vec{b},\vec{a}}$ from density matrix renormalization group (DMRG). The correlation matrix includes all Pauli correlators of weight $k \leq 2$ with maximum range $r_M = 4$, and its dimension scales approximately as $\sim L$. For illustration, we show results for a system of size \(L = 64\), using $N = 16000$ snapshots. We observe generally good agreement between the learned and exact matrices, with maximum deviations on the order of $\sim 0.03$ appearing primarily in higher-weight correlators.
In Figs.~\ref{fig:M} (c) and (d), we further examine select two-point functions, specifically \(\langle Z_0 Z_r \rangle\) and \(\langle X_0 X_r \rangle\), respectively. Here as well we find good agreement with results obtained from DMRG.

In Figs.~\ref{fig:Scaling} (a), (b) and (c), we show how the error in the energy density $\left( \hat{H} - \avg{H} \right)/L$ and the standard deviation in computing \emph{contiguous} (operating non-trivially only on contiguous sites) Pauli operators of weight $k = 2$, $k = 5$ scales with the number of snapshots $N$ and system size $L$. 

The inset shows the same data but plotted as a function of system size $L$. 
The weight $k = 2$ Pauli operators, which also include those in the Hamiltonian, tend to have a slightly larger average error as compared to the energy density; see Fig.~\ref{fig:Scaling}. We note that, in an idealized shadow tomography experiment, estimating a weight $k$ operator to accuracy $\epsilon$ requires $N = \mathcal{O} \left( \frac{3^k}{\epsilon^2} \right)$ snapshots. Setting $\epsilon = 0.02, k = 2$ would imply $N = 22,500$ snapshots are needed to obtain such accuracy, whereas only $N = 2000$ snapshots are needed in practice, see Fig.~\ref{fig:Scaling}. We note that since we are not truly performing a tomography experiment, and instead fine tuning the snapshots in a variational procedure, the accuracy obtained at a given $N$ may be even better. 

Furthermore, the errors in panels $(a)$ and $(b)$ show little dependence on $N$, suggesting that convergence is reached at a relatively small number of snapshots for correlators directly included in the PSD constraint. For such correlators, the error is likely dominated by the choice of imposed PSD constraints rather than finite $N$ fluctuations. The error is largely independent of system size $L$, in agreement with expectations from classical shadows. 

In Fig.~\ref{fig:Scaling}(c), we study errors associated with contiguous Pauli operators of weight $k = 5$, which do not appear in the imposed PSD constraints. Remarkably, the observed accuracy follows the classical-shadow prediction (red curve) almost perfectly, despite these operators not being included directly in the optimization cost function. While only the average error is shown here, the full distribution of observed errors further supports the conclusion that the optimized ensemble of snapshots faithfully reproduces higher-weight correlations beyond the subset explicitly constrained during optimization; for more details, see SM \cite{footnote1}. This demonstrates that the snapshot-based representation captures many-body correlations more comprehensively than standard RDMAs. 

We briefly discuss the scaling properties of the algorithm. As noted above, we expect the number of snapshots needed for accurate computation of the correlators to scale only logarithmically with system size $L$; this is borne out in the numerical data, at least for the system sizes studied. Moreover, the computation of correlators from snapshots can be parallelized. For the correlation matrix constructed from finite weight and range Pauli operators, each optimization epoch requires the computation of $\mathcal{O} (L^2)$ correlators. This can be further reduced to $\mathcal{O} (L)$ computations if translational invariance is employed. The step dominating the computational cost at large $L$ is the computation of the inverse of $M$ needed as part of the gradient calculation of the PSD cost---for a matrix of dimension $\mathcal{O} \left(L \times L\right)$, such computation scales as $\mathcal{O} \left( L^3 \right)$. Although we do not implement it, using translational invariance, the constraint matrix $M$ can be broken up into  $\mathcal{O} (L)$ block diagonal chunks of shape $\mathcal{O} (1) \times \mathcal{O}(1) $ corresponding to different momenta, reducing the complexity of the algorithm further. 

\textbf{\textit{Summary and Future directions.---}}We have introduced a novel variational approach for computing ground states of quantum many-body systems by representing them via distributions over measurement outcomes or snapshots. 
The method draws inspiration from reduced density matrix approaches which perform a constrained search in the space of correlators of low weight observables. But unlike reduced matrix approaches, the representation in terms of snapshots allows us to predict higher weight observables that are not employed in the constraints. We find that the accuracy of predicting such correlators improves systematically with increasing snapshot count $N$, and in close agreement with the scaling predicted by classical shadows. These observations suggest that enforcing exact PSD constraints on a restricted subset of physically relevant Pauli correlators can be sufficient to generate an ensemble of snapshots that accurately predicts ground state energy and a wide range of observables. 

Several directions for refinement and extension remain open. From a theoretical perspective, a systematic study of the relationship between the set of correlators included in the PSD constraint and the corresponding set of realizable quantum states would be valuable. Such an analysis could help identify constraint sets that minimize computational cost while maintaining a desired level of accuracy. On the algorithmic side, incorporating translational and internal symmetries directly into the optimization procedure could substantially improve scalability. The present implementation also does not explicitly enforce purity, which may become relevant in systems with degenerate ground states and may help further improve the accuracy of the studied gapless Hamiltonians, see SM \cite{footnote1}.\\ 

The representation of the state in terms of measurement outcomes is quite general, and can therefore capture states with different entanglement patterns, as well as represent states in higher dimensions. Investigation of the method in higher dimensional systems may be particularly interesting given that conventional methods such as tensor networks are more limited. Furthermore, the representation is naturally amenable to describing mixed states, and concomitantly, dynamics under a bath which provides dephasing and/or relaxation. This presents another promising direction for future work.

\textbf{\textit{Acknowledgments.---}}
PR acknowledges funding support from NSERC. KA acknowledges support from the US Department of Energy, Office of Science, Basic Energy Sciences. KA and PR acknowledge HPC facilities at Argonne National Laboratory for computational resources that were used to obtain results discussed in this manuscript.  
\begin{widetext}
\section{Classical shadow parametrization}

\subsubsection{Classical shadow preliminaries}
In the manuscript, the measurement outcomes which parametrize the variational quantum state are  given by
\begin{align}
\ket{b_l} (\vec{\theta}_l) = \bigotimes_{j = 1}^L e^{i\frac{1}{2}\theta_{j,l}\sigma^x_j}\ket{0} \; \;,  \; \; \mathcal{U}_l = \bigotimes_{j = 1}^L U_{j,l}, 
\end{align}
where $\ket{b_l}$ represents a possible measurement outcome obtained after the putative application of a random unitary $\mathcal{U}_l$. This differs from the conventional classical shadow protocol which has discrete measurement outcomes $\ket{b_l} \in \{0,1\}^{\otimes L}$. In this section, we show how to arrive at the inverted snapshots
\begin{align}
    \hat{\rho}_l = \bigotimes_{j = 1}^L \left(1 + 3 \vec{n}_{j,l} (\theta_{j,l} ) \cdot \vec{\sigma}_j \right)
\end{align}
First, we briefly recall the single-site twirling classical shadow protocol. Given an unknown state $\rho$ to be learned, an experimentalist repeatedly applies a tensor product of single-site unitaries $\mathcal{U}_l = \otimes_{j=1}^LU_{j,l}$ where $l$ labels the $l^{\text{th}}$ experiment and $j$ labels the sites. For each pair of site and experiment $(j,l)$ the unitary $U_{j,l}$ is sampled randomly according to the Haar random measure on a single qubit. After each application of the unitary $\mathcal{U}_l$ each qubit $j$ is measured in the $Z$ basis, resulting in strings of measurement outcomes $\ket{b_l} \in \{0,1\}^{\otimes L}$ (with the convention $Z\ket{0} = \ket{0}$ and $Z\ket{1} = -\ket{1}$). One may then consider the average 
\begin{align}
A_N = \frac{1}{N}\sum_{l=1}^{N}\mathcal{U}_{l}^{\dagger}\ket{b_l}\bra{b_l}\mathcal{U}_{l}
\end{align}
with the goal of finding a relationship between $A_N$ and the unknown quantum state $\rho$.
In the limit of large $N$ the operator $A_N$ converges to 
\begin{align}
\lim_{N \rightarrow \infty}A_N = \frac{1}{2^L}\sum_{b \in \{0,1\}^{\otimes L}}\int \bra{b}\mathcal{U}\rho\mathcal{U}^{\dagger}\ket{b}\mathcal{U}^{\dagger} \ket{b}\bra{b} \mathcal{U} d\mathcal{U} \equiv A_{\infty}
\end{align}
where $d\mathcal{U} = \otimes_{j=1}^L dU_j$ is the tensor product of Haar random measures over each qubit, $\bra{b}\mathcal{U}\rho\mathcal{U}^{\dagger}\ket{b}$ is the probability to obtain $\ket{b}$ given the state $\rho$ and the applied unitary $\mathcal{U}$. $A_{\infty}$ can be viewed as a linear, trace and positivity preserving operation on the unknown state $\rho$ and can be interpreted as a quantum channel. As such we consider the quantum channel $\mathcal{M}$ defined by its action on an operator $O$
\begin{align}\label{eq:shadow_channel}
\mathcal{M}(O) = \frac{1}{2^L}\sum_{b \in \{0,1\}^{\otimes L}}\int \bra{b}\mathcal{U}O\mathcal{U}^{\dagger}\ket{b}\mathcal{U}^{\dagger} \ket{b}\bra{b} \mathcal{U} d\mathcal{U}.
\end{align}
In the particular case of local Haar random unitaries considered here, the channel $\mathcal{M}$ is invertible. It thus holds by construction that
\begin{align}
\mathcal{M}^{-1}(A_{\infty}) = \mathcal{M}^{-1}(\mathcal{M}(\rho)) = \rho
\end{align}
and it holds true due to the linearity of quantum channels that 
\begin{align}
\lim_{N\rightarrow \infty}\mathcal{M}^{-1}(A_{N}) = \rho.
\end{align}
As such, the quantity $\mathcal{M}^{-1}(A_{N})$ provides an unbiased empirical estimate of the unknown quantum state $\rho$
\begin{align}
\rho_N = \frac{1}{N}\sum_{l=1}^{N}\mathcal{M}^{-1}\left(\mathcal{U}_{l}^{\dagger}\ket{b_l}\bra{b_l}\mathcal{U}_{l}\right) \;\;\; \text{with }\lim_{N \rightarrow \infty} \rho_N = \rho
\end{align}
An analysis of equation \ref{eq:shadow_channel} reveals that Pauli strings $P_{\vec{a}}$ are eigenmodes of the channel, and for single-site twirling the channel further decomposes to a tensor product of local quantum channels such that 
\begin{align}
\mathcal{M}(P_{\vec{a}}) = \bigotimes_{j=1}^{L}\mathcal{M}_j(P_{a_j})
\end{align}
where $\mathcal{M}_j(I_j) = I_j$, $\mathcal{M}_j(X_j) = (1/3)X_j$, $\mathcal{M}_j(Y_j) = (1/3)Y_j$ and $\mathcal{M}_j(Z_j) = (1/3)Z_j$. Consequently the inverse is such that $\mathcal{M}_j^{-1}(I_j) = I_j$, $\mathcal{M}_j^{-1}(X_j) = 3X_j$, $\mathcal{M}_j^{-1}(Y_j) = 3Y_j$ and $\mathcal{M}_j^{-1}(Z_j) = 3Z_j$. Using these facts leads to the simplified expression for $\rho_N$ given by 
\begin{align}
\rho_N = \frac{1}{N}\sum_{l=1}^N\bigotimes_{j=1}^{L}\mathcal{M}_j^{-1}(U_{j,l}^{\dagger}\ket{b_l}_j\bra{b_l}_jU_{j,l}).
\end{align}
From there we may write

\begin{align}
U_{j,l}^{\dagger}\ket{b_l}_j\bra{b_l}_jU_{j,l} = \frac{1}{2}\left(I_j + (n_{j,l})_x X + (n_{j,l})_yY + (n_{j,l})_z Z \right)
\end{align}
with $(n_{j,l})_a$ given by 
\begin{align}
(n_{j,l})_a = \frac{1}{2}\operatorname{Tr}\left(U_{j,l}^{\dagger}\ket{b_l}_j\bra{b_l}_jU_{j,l} \sigma^{a}\right)
\end{align}
from which we directly read off the snapshots 
\begin{align}
\hat{\rho}_l =  \frac{1}{2^L}\bigotimes_{j=1}^L (I_j + 3 \vec{n}_{j,l}\cdot \vec{\sigma})
\end{align}
such that 
\begin{align}
\rho_N = \frac{1}{N}\sum_{l=1}^N \hat{\rho}_l
\end{align}
with $\vec{\sigma} = (X,Y,Z)$ and $\vec{n}_{j,l} = ((n_{j,l})_x,(n_{j,l})_y,(n_{j,l})_z)$. To evaluate the effectiveness of the parametrization for estimating various Pauli sting observables, one then considers the estimators
\begin{align}
\operatorname{Tr}\left(\rho_N P_{\vec{a}}\right) = \frac{1}{N}\sum_{l=1}^N \operatorname{Tr}(\hat{\rho}_lP_{\vec{a}}).
\end{align}
The $\hat{\rho}_l$ are statistically independent, and thus the variance of $\operatorname{Tr}(\hat{\rho}_lP_{\vec{a}})$ controls the variance of the sum by virtue of the central limit theorem. One finds that
\begin{align}
\mathbb{E}(\operatorname{Tr}\left(\rho_N P_{\vec{a}}\right)^2) = \frac{1}{2^L}\sum_{b \in \{0,1\}^{\otimes L}}\int \bra{b}\mathcal{U}\mathcal{M}^{-1}(P_{\vec{a}})\mathcal{U}^{\dagger}\ket{b}^2 d\mathcal{U} = 3^{w(P_{\vec{a}})} \geq \operatorname{Var}(\operatorname{Tr}\left(\rho_N P_{\vec{a}}\right)^2)
\end{align}
where $w(P_{\vec{a}})$ is the total number of non-identity Pauli operators in the Pauli string $P_{\vec{a}}$. Remarkably this is completely independent of the unknown state $\rho$. As such, given a sufficiently large set of $N$ independent samples, any Pauli operator  $P_{\vec{a}}$ can be estimated with a corresponding standard deviation $\Delta_{\vec{a},N} = \sqrt{3^{w(P_{\vec{a}})}/N}$ 

\subsubsection{Alternate parametrization}
We now consider the alternate parametrization of the measurement outcomes used in the manuscript. To do so, let us consider the additional application of a random unitary $R_{x,j,l} = e^{i \frac{1}{2}\alpha_{j,l} \sigma_j^x}$ where the angles $\alpha_{j,l}$ are sampled from a distribution that is uncorrelated with that of the Haar random unitaries. With this assumption, the operator $A_N$ is deformed to 
\begin{align}
A_N = \frac{1}{N}\sum_{l=1}^{N}\mathcal{U}_{l}^{\dagger}\mathcal{R}_{x,l}^{\dagger}\ket{b_l}\bra{b_l}\mathcal{R}_{x,l}\mathcal{U}_{l}
\end{align}
where we used
\begin{align}
\mathcal{R}_{x,l} = \bigotimes_{j=1}^LR_{x,j,l}.
\end{align}
Following a similar procedure as before, we arrive that the channel
\begin{align}
\mathcal{M}_{R}(O) = \frac{1}{2^L}\sum_{b \in \{0,1\}^{\otimes L}}\int\int \bra{b}\mathcal{R}_{x,l}\mathcal{U}O\mathcal{U}^{\dagger}\mathcal{R}_{x,l}^{\dagger}\ket{b}\mathcal{U}^{\dagger} \mathcal{R}_{x,l}^{\dagger}\ket{b}\bra{b}\mathcal{R}_{x,l} \mathcal{U} d\mathcal{U}d\vec{\alpha}.
\end{align}
with $d\vec{\alpha} = \prod_{j=1}^L d\alpha_j/\pi$. From here, we make use of the fact that the Haar random average is invariant under unitary multiplications. As such, the matrices $\mathcal{R}_{x,j,l}^{\dagger}$ can be absorbed in the Haar random unitaries $U_{j,l}$ and the integration over the angles $\alpha_{i,j}$ can be performed trivially to arrive at the conclusion that the channel remains unchanged $\mathcal{M}_{R} = \mathcal{M}$. This leads to the modified unbiased estimator 
\begin{align}
\rho_{N,R} = \frac{1}{N}\sum_{l=1}^N\bigotimes_{j=1}^{L}\mathcal{M}_j^{-1}(U_{j,l}^{\dagger}R_{x,j,l}^{\dagger}\ket{b_l}_j\bra{b_l}_j\mathcal{R}_{x,j,l}U_{j,l})
\end{align}
where $\rho_{N,R}$ converges to $\rho$ in the large $N$ limit. Expanding into a tensor product one arrives at 
\begin{align}
&\rho_{N,R} = \frac{1}{N}\sum_{l=1}^N\bigotimes_{j=1}^{L}\mathcal{M}_j^{-1}(U_{j,l}^{\dagger}e^{-i\frac{1}{2}(b_{l,j} + (1 - b_{l,j})\pi + \alpha_{j,l})X}\ket{0}_j\bra{0}_je^{i\frac{1}{2}(b_{l,j} + (1 - b_{l,j})\pi + \alpha_{j,l})X}U_{j,l}) \nonumber \\
& = \frac{1}{N}\sum_{l=1}^N\bigotimes_{j=1}^{L}\mathcal{M}_j^{-1}(U_{j,l}^{\dagger}e^{-i\frac{1}{2}\theta_{j,l}X}\ket{0}_j\bra{0}_je^{i\frac{1}{2}\theta_{j,l}X}U_{j,l})
\end{align}
where we have used the substitution 
\begin{align}
\theta_{j,l} = b_{l,j} + (1 - b_{l,j})\pi + \alpha_{j,l}.
\end{align}
Since the variable $\alpha_{j,l}$ is continuous (in contrast to the $b_{j,l}$), it follows that the parameters $\theta_{j,l}$ are also continuous variables, which are treated as the optimization parameters in the manuscript. The modified snapshots are then given by
\begin{align}
\hat{\rho}_{l} =  \frac{1}{2^L}\bigotimes_{j=1}^L (I_j + 3 \vec{n}_{j,l}(\theta_{j,l})\cdot \vec{\sigma}) 
\label{eq:invertedrho}
\end{align}
where the explicit expression of $\vec{n}_{j,l}(\theta_{j,l})$ in terms of $\theta_{j,l}$ and the applied unitary $U_{j,l}$ is discussed in the following sections.
\subsubsection{Statistical independence of the $\alpha_{j,l}$}
The derivation of $\hat{\rho}_l$ formally assumes that the optimized angles $\alpha_{j,l}$ remain statistically independent from the sampled Haar-random unitaries. 

Significant deviations from statistical independence would indicate that the optimized variational distribution effectively realizes inversion statistics distinct from those of standard classical shadows. Such deviations could, for example, modify the expected fluctuation amplitudes associated with the estimation of certain observables, or worse, lead to biased estimators. It is therefore relevant to investigate whether the optimized variational angles converge toward a distribution that approximately preserves statistical independence from the sampled Haar-random measurement directions. \\

To investigate this question, we study the correlation between the Bloch-sphere vectors associated with the sampled Haar-random unitaries and those generated by the optimized variational angles. For each snapshot indexed by $(j,l)$, the Haar-random unitary defines a Bloch vector
\[
\mathbf{v}^{\mathrm{Haar}}_{j,l}
=
\begin{pmatrix}
\sin\alpha_{j,l}\cos\phi_{j,l} \\
\sin\alpha_{j,l}\sin\phi_{j,l} \\
\cos\alpha_{j,l}
\end{pmatrix},
\]
where $(\alpha_{j,l},\phi_{j,l})$ denote the polar and azimuthal angles parametrizing the Haar-random measurement direction.

As for the optimized variational angle $\theta_{j,l}$ we consider the map
\[
\mathbf{v}^{\mathrm{opt}}_{j,l}(n)
=
\begin{pmatrix}
0 \\
-\sin(n\theta_{j,l}) \\
\cos(n\theta_{j,l})
\end{pmatrix},
\]
where $n$ is an integer controlling the angular winding of the variational trajectory on the Bloch sphere.

An important distinction arises between the cases $n=1$ and $n=2$. Taking $n=2$ effectively removes the contribution of the measurement outcomes $b_j$ from the distribution of the optimized angles $\theta_{j,l}$. Consequently, if the optimization procedure preserves approximate statistical independence from the Haar-random unitaries, we expect the correlation between $\mathbf{v}^{\mathrm{opt}}_{j,l}(n=2)$ and $\mathbf{v}^{\mathrm{Haar}}_{j,l}$ to remain approximately constant throughout optimization. In contrast, the vectors $\mathbf{v}^{\mathrm{opt}}_{j,l}(n=1)$ are expected to develop increasing correlations with optimization epoch $t$, since the measurement outcomes themselves become correlated with the applied random unitaries during training.

To quantify these effects, we consider the ensembles
\[
\left\{
\mathbf{v}^{\mathrm{Haar}}_{j,l}
\right\},
\qquad
\left\{
\mathbf{v}^{\mathrm{opt}}_{j,l}(n)
\right\},
\]
and compute their component-wise cross-correlation matrix. Prior to computing correlations, each Cartesian component is centered and normalized across snapshots according to
\[
\tilde v_{j,l,\alpha}
=
\frac{
v_{j,l,\alpha}
-
\langle v_{j,l,\alpha} \rangle
}{
\sigma(v_{j,l,\alpha})
},
\]
where $\alpha\in\{x,y,z\}$, $\langle\cdot\rangle$ denotes the average over snapshots, and $\sigma(v_{j,l,\alpha})$ is the corresponding standard deviation.

Centering removes trivial directional offsets associated with nonzero mean Bloch vectors, while normalization prevents the correlation metric from being dominated by Cartesian components with larger variations across snapshots. The resulting correlation matrix is defined as
\[
C_{j,\alpha,\beta}(n)
=
\frac{1}{N}
\sum_{l=1}^N
\tilde v^{\mathrm{opt}}_{j,l,\alpha}(n)
\tilde v^{\mathrm{Haar}}_{j,l,\beta}.
\]
The matrix element $C_{j,\alpha,\beta}(n)$ therefore quantifies the linear correlation between the $\alpha$-component of the optimized Bloch vector and the $\beta$-component of the Haar-random Bloch vector at site $j$.

As a scalar measure of the total correlation strength, we consider the squared Frobenius norm
\[
\|C_j\|_F^2(n)
=
\sum_{\alpha,\beta}
C_{j,\alpha,\beta}(n)^2.
\]
This quantity is invariant under redistributions of correlation between Cartesian components and provides a compact measure of the total linear dependence between the optimized and Haar-random measurement directions. Values close to zero indicate weak component-wise correlations and therefore approximate statistical independence, whereas larger values indicate that the optimized variational angles retain significant information about the original Haar-random unitaries.

We study the correlations at site $j = 0$ (we do not find statistically significant dependence of the results on the chosen site $j$), see Fig. \ref{fig:haar_correlations}, which shows that $C_0(1)$ increases steadily whilst $C_0(2)$ remains approximately constant, suggesting that the optimized solution produces an ensemble of snapshots that is consistent with the classical shadow framework. This is further corroborated by the almost exact agreement of the average error of weight $k = 5$ Pauli operators with classical shadow prediction discussed in the main text.

\begin{figure}
    \includegraphics[width=0.5\textwidth]{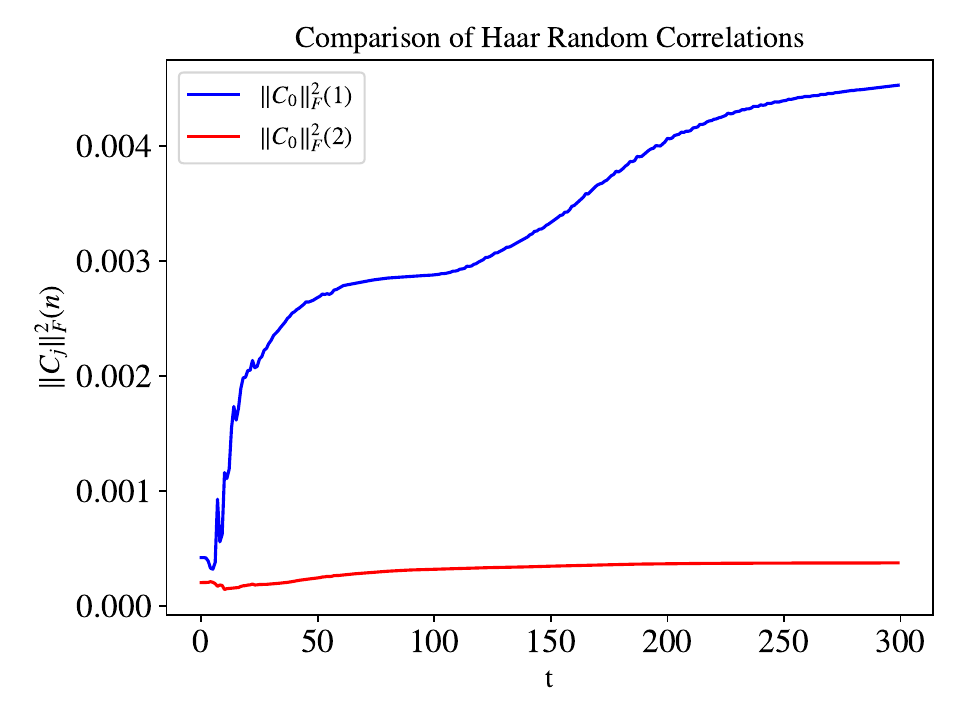} 
\caption{Frobenius norm $\|C_j\|_F^2(n)$ for $n = 1,2$ at site $j = 0$.}\label{fig:haar_correlations}
\end{figure}
\section{Other models}
To provide a more complete picture, we study the performance of the proposed variational algorithm for other Hamiltonians, whose results are displayed in Fig. \ref{fig:othermodels}. Respectively, we consider the models

\begin{align}
\text{Likely Gapless}: \nonumber \\
&H_1 = \sum_j0.25\, Z_j Z_{j+1}
+ 0.3\, X_j X_{j+1}
+ 0.3\, Y_j Y_{j+1}
+ 0.25\, Z_j
+ 0.3\, X_j \nonumber \\
\text{Gapped}: \nonumber \\
&H_2 = \sum_j 2X_j - Z_jZ_{j+1} \nonumber \\ 
&H_3 = \sum_j 0.12Z_jZ_{j+1} + 0.25X_jX_{j+1} + 0.25Y_jY_{j+1} - 2Z_j \nonumber \\
\text{Gapless}: \nonumber \\
&H_4 = \sum_j X_j - Z_jZ_{j+1} \nonumber \\ 
\end{align}

with periodic boundary conditions.

\begin{table}
\centering
\begin{tabular}{c c c c c}
\hline
Model & Internal symmetry & Time reversal & Parity & Phase \\
\hline
$H_1$ & None & Yes & Yes & Likely gapless from DMRG data \\
$H_2$ & $\mathbb{Z}_2$, generated by $\prod_j X_j$ & Yes & Yes & Gapped \\
$H_3$ & $U(1)$, generated by $\sum_j Z_j$ & Yes & Yes & Gapped \\
$H_4$ & $\mathbb{Z}_2$, generated by $\prod_j X_j$ & Yes & Yes & Gapless \\
\hline
\end{tabular}
\caption{Summary of the symmetries and spectral properties of the benchmark Hamiltonians.}
\label{tab:benchmark_hamiltonians}
\end{table}

The properties of the model are summarized in \ref{tab:benchmark_hamiltonians}. The obtained optimized correlators are displayed in Fig.~\ref{fig:othermodels}. 

We find that the gapless models $H_1$ and $H_4$ appear to exhibit the largest deviation between the calculated and exact correlators. We believe this behavior originates from  how strongly local positivity conditions restrain long-range expectation values in a given model. More precisely, once a set of low-range expectation values satisfying the imposed PSD constraints has been fixed, there generally exists a family of compatible long-range correlators that do not modify either the local PSD constraints and that do not appear in the energy expectation value $\langle H \rangle$. Consequently, during optimization, the low-range correlators are strongly constrained by the combined requirements of minimizing $\langle H \rangle$ while maintaining positivity, whereas sufficiently long-range correlators can admit any of the compatible choice. 
This naturally introduces an entropic bias toward smaller long-range Pauli expectation values. Indeed, decreasing the magnitude of unconstrained correlators drives the reconstructed state closer to the maximally mixed state $\rho = I/D$, which tends to reduce the PSD cost. This can be understood from the relation
\[
\mathrm{Tr}(M M^\dagger)
=
\sum_{P\in S} \langle P\rangle^2
=
\sum_i \lambda_i^2,
\]
where $S$ denotes the set of Pauli operators entering the matrix $M$, and $\lambda_i$ are its eigenvalues. Reducing unconstrained expectation values therefore decreases the overall spectral weight of the PSD matrices, reducing the $\log\det$ cost. As a result, long-range correlators that are not directly constrained by the optimization tend to decay toward zero. 
This interpretation is consistent with the observed discrepancies, which systematically correspond to expectation values whose amplitudes are underestimated relative to the exact result. We find this effect to be particularly pronounced for the critical transverse-field Ising model, where long-range correlations are expected to play a dominant role. Interestingly, in this regime, increasing the number of snapshots can further suppress long-range correlators. This seemingly counterintuitive behavior is nevertheless consistent with the above interpretation: the additional variational degrees of freedom can be used to further reduce the PSD cost while leaving the energy approximately unchanged. \\
\begin{figure}
    \includegraphics[width=1.0\textwidth]{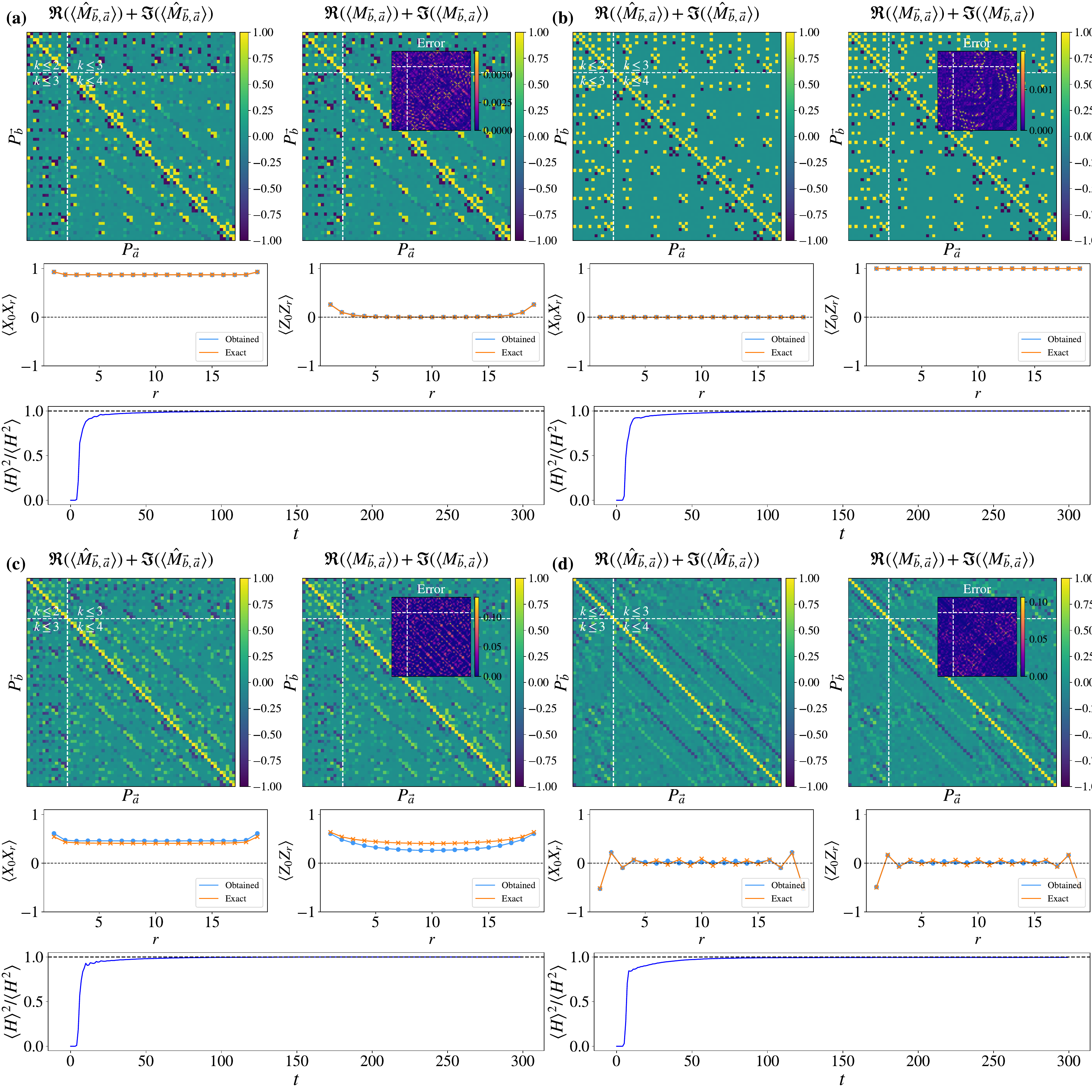} 
    \caption{
 Exact and approximate matrix $\hat{M}$ obtained for (a) $H = \sum_j X_j - 2Z_jZ_{j+1}$, (b) $H = \sum_j 0.12Z_jZ_{j+1} + 0.25X_jX_{j+1} + 0.25Y_jY_{j+1} - 2Z_j$ (c) $H = \sum_j X_j - Z_jZ_{j+1}$ and (d) $\sum_j0.25\, Z_j Z_{j+1} + 0.3\, X_jX_{j+1}+ 0.3\, Y_j Y_{j+1}+ 0.25\,Z_j + 0.3\, X_j$ for $L = 20$, $N = 16000$, $r_M = 2$. The top panels showcase the correlation matrix obtained over a subset of $4$ contiguous sites. The figures displayed in the middle panels show the values of the correlators $Z_0Z_r$ and $X_0X_r$. The bottom panels show the ratio $\langle H\rangle^2/\langle H^2\rangle$ as a function of the optimization epoch $t$.}\label{fig:othermodels}
\end{figure}
\section{Predictions beyond $k \leq 4$ Pauli operators}
Despite the fact that the cost function only involves Pauli operators of weight $k \leq 4$, it is possible to obtain estimates of the expectation value of Pauli operators with a larger weight due to snapshot parameterization of the optimized state. This is in contrast to more conventional RDMA's methods \cite{MAZZIOTTI2002, MAZZIOTTI2005, 
HAMMOND2005,
MAZZIOTTI2005_spin,
MAZZIOTTI2006, ALEXANDER2006,
MAZZIOTTI2007, 
MAZZIOTTI2010, MAZZIOTTI2014,  MAZZIOTTI2020, MAZZIOTTI2024} whose predictive capabilities are limited to the parametrized operators that appear in the cost function. Remarkably, despite the lack of constraints involving the larger-weight Pauli operators, we find that the obtained estimates agree well with the exact values; see Fig. \ref{fig:other_models_k=5}. In the top plots, we show the exact expectation values for all $k = 5$ contiguous Pauli operators, and below we show the predicted value of the same operators. The number of $k = 5$  weight operators with sizable expectation values can be small (depending on the model), and the reader may worry that our claim that the \emph{average} error between prediction and exact result is simply due to this fact. These results illustrate that the protocol clearly captures accurately the expectation values of high-weight operators irrespective of the magnitude of the expectation value. 
This suggests that the predictive power of the developed method goes beyond the operators involved in the cost function. 

\begin{figure}[htbp]
    \centering
    \includegraphics[width=1\textwidth,trim=9cm 0cm 8cm 0cm]{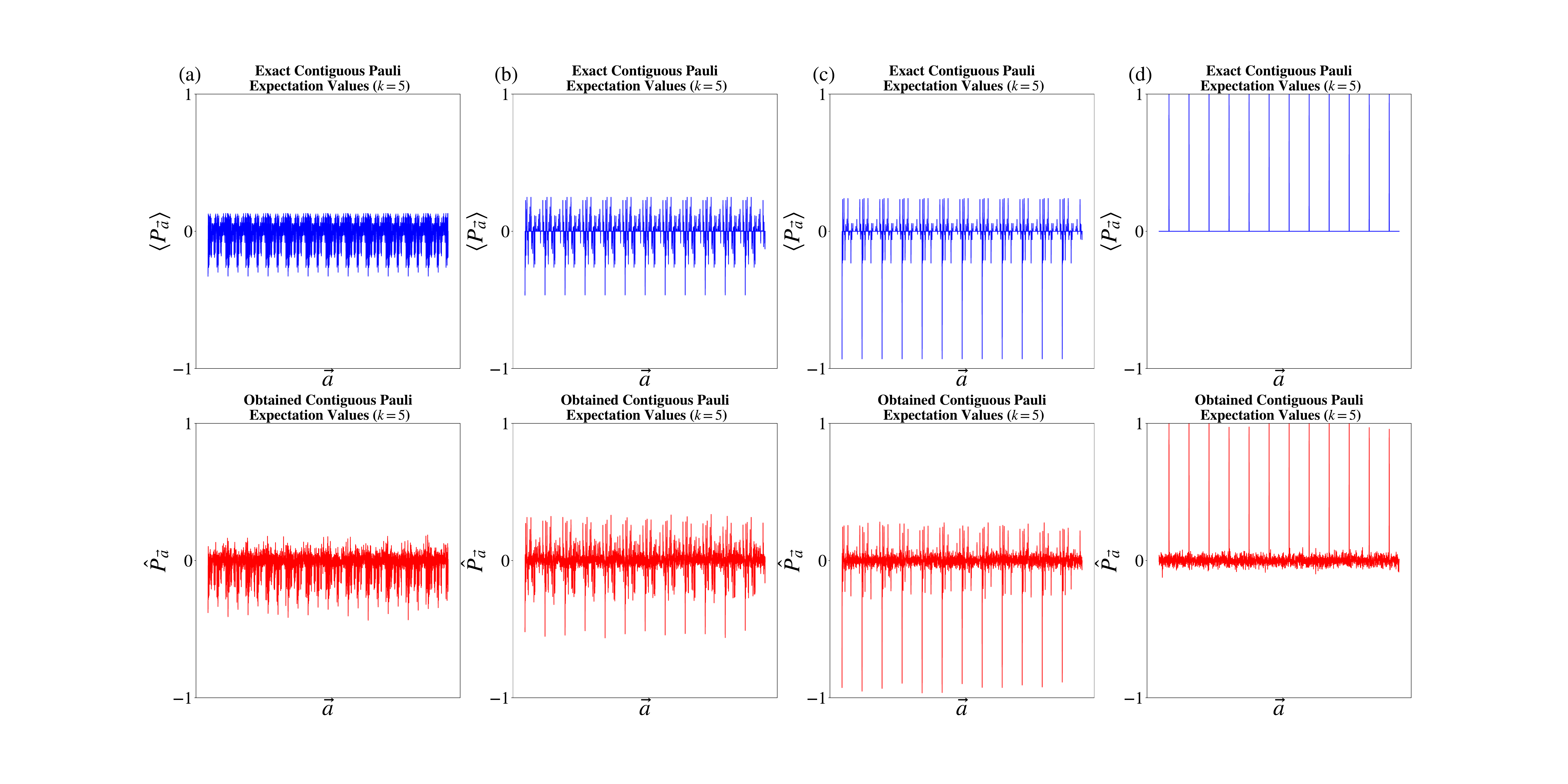}  
    \caption{ (a) $\sum_j0.25\, Z_j Z_{j+1} + 0.3\, X_jX_{j+1}+ 0.3\, Y_j Y_{j+1}+ 0.25\,Z_j + 0.3\, X_j$  (b) $H = \sum_j X_j - Z_jZ_{j+1}$, (c) $H = \sum_j 2X_j - Z_jZ_{j+1}$ and (d) $H = \sum_j 0.12Z_jZ_{j+1} + 0.25X_jX_{j+1} + 0.25Y_jY_{j+1} - 2Z_j$ for $L = 12$, $N = 256000$, $r_M = 2$. The top panels display the exact expectation value of all contiguous Pauli operator $\langle P_{\vec{a}} \rangle$ of weight $k = 5$. The bottom panels display the corresponding estimate from the optimized snapshots.}
    \label{fig:other_models_k=5}.
\end{figure}

\section{Snapshots generation}
We start by discussing the calculation of snapshots $\hat{\rho}_l$. First, as noted in the main text, the shadow tomography prescription for the density matrix estimate obtained by inverting a particular snapshot $l$ is given by $    \hat{\rho}_l = \frac{1}{2^L}\bigotimes_{j = 1}^L \left(I + 3 \vec{n}_{j,l} (\theta_{j,l} ) \cdot \vec{\sigma}_j \right) $. Here $\vec{n}_{j,l}$ is a unit vector whose components are proportional to the expectation value of the operator $X/Y/Z$ at the site $j$ in the state obtained by applying the inverse unitary $U^\dagger_{j,l}$ to the measured state. For example, if snapshot $l$ is obtained applying $U_l = \prod_j U_{j,l}$ to $\rho$, and the measurement result gives the state $e^{-i \frac{1}{2} \theta_{j,l} X} \ket{0}$, then $(\vec{n}_{j,l})_{X/Y/Z} \propto \bra{0}e^{i\frac{1}{2}\theta_{j,l}X}U_{j,l} \cdot (X/Y/Z) \cdot U_{j,l}^{\dagger}e^{-i\frac{1}{2}\theta_{j,l}X}\ket{0}$. The Haar random unitaries $U_{j,l}$ \cite{Zyczkowski1994} can be parametrized using three angles $\phi_{j,l}, \omega_{j,l}, \alpha_{j,l}$ per qubit and snapshot as

\begin{align}
&U_{j,l} (\phi_{j,l},\omega_{j,l}, \alpha_{j,l}) =
\begin{pmatrix}
e^{-i(\phi_{j,l} + \omega_{j,l})/2} \cos(\alpha_{j,l}/2) & -e^{i(\phi_{j,l} - \omega_{j,l})/2} \sin(\alpha_{j,l}/2) \\
e^{-i(\phi_{j,l} - \omega_{j,l})/2} \sin(\alpha_{j,l}/2) & e^{i(\phi_{j,l} + \omega_{j,l})/2} \cos(\alpha_{j,l}/2)
\end{pmatrix}.
\end{align}
Conjugating the Pauli operator $\sigma^a$ with $U_{i,l}$ yields
\begin{align}
U_{j,l}\sigma^{a}U_{j,l}^{\dagger} = \sum_{b \in 
 {I,X,Y,Z}}u_{b,a,j,l}\sigma^b
\end{align}
from which one can deduce
\begin{align}
(\vec{n}_{j,l})_a = 
u_{I,a,j,l} -\sin(\theta_{j,l})u_{Y,a,j,l} + \cos(\theta_{j,l} )u_{Z,a,j,l}.
\end{align}
The parameters $u_{b,a,j,l}$ are explicitly given by
\begin{align}
&u_{I,a,j,l} = \delta_{I,a}\nonumber \\
&u_{Y,X,j,l} = (\cos(\omega_{j,l})\sin(\phi_{j,l}) + \cos(\alpha_{j,l})\cos(\phi_{j,l})\sin(\omega_{j,l}))\nonumber \\
&u_{Z,X,j,l} = -\cos(\phi_{j,l})\sin(\alpha_{j,l})\nonumber \\
&u_{Y,Y,j,l} = (\cos(\phi_{j,l})\cos(\omega_{j,l}) - \cos(\alpha_{j,l})\sin(\phi_{j,l})\sin(\omega_{j,l}))\nonumber \\
&u_{Z,Y,j,l} = \sin(\alpha_{j,l})\sin(\phi_{j,l})\nonumber \\
&u_{Y,Z,j,l} = \sin(\alpha_{j,l})\sin(\omega_{j,l})\nonumber \\
&u_{Z,Z,j,l} = \cos(\alpha_{j,l}).
\end{align}
These parameters are randomly selected at the beginning of the algorithm for all pairs $(j,l)$, $l=1,...,N$,  $j=1,...,L$ and remain constant throughout the optimization. More explicitly, for every pair $(j,l)$, a random triplet $(\phi_{j,l},\omega_{j,l}, \alpha_{j,l}) = (\pi\gamma_0,\pi\gamma_1, \arccos (\gamma_3))$ is generated, where the $\gamma_n$ are selected at random in the interval $[-1,1]$, ensuring that the Haar random unitaries are sampled uniformly. 
\section{Calculation of observables}
The only ingredient necessary for the calculation of the expectation value of any operator $\hat{O}$ in this framework is the expectation value of a Pauli operator $P_{\vec{a}} = \bigotimes_{j=1}^L \sigma^{a_j} $ where $\vec{a}$ is a vector whose elements (either $I,X,Y,Z$) specify the Pauli operator at site $j$ appearing in the tensor product.
The expectation value $\operatorname{Tr}(\hat{\rho}_lP_{\vec{a}})$ may then be written as
\begin{align}\label{eq:pauliprod}
\operatorname{Tr}(\hat{\rho}_lP_{\vec{a}}) = \prod_{j=1}^L 3^{w(P_{\vec{a}})}(\vec{n}_{j,l})_{a_j}.
\end{align}
Using equation \ref{eq:pauliprod}, it is straightforward to compute $\hat{M}$ and $\hat{H}$, which are given explicitly by
\begin{align}
\hat{M}_{\vec{b},\vec{a}} = \frac{1}{N}\sum_{l=1}^N\operatorname{Tr}(\hat{\rho}_l P_{\vec{b}}P_{\vec{a}}), \quad 
\hat{H} = \frac{1}{N}\sum_{l=1}^N\sum_{j=1}^L\operatorname{Tr}(\hat{\rho}_lh_j).
\end{align}
\section{Gradient calculation}
The derivative of any observable $\hat{O}$ may be extracted from the derivative of $\operatorname{Tr}(\hat{\rho}_lP_{\vec{a}})$, which is given by
\begin{align}
\frac{\partial}{\partial \theta_{j,l}}\operatorname{Tr}(\hat{\rho}_lP_{\vec{a}}) = 
\left(\frac{\partial}{\partial \theta_{j,l}} (\vec{n}_{j,l})_{a_j}\right)\prod_{p\neq j}^L 3^{w(P_{\vec{a}})}(\vec{n}_{p,l})_{a_p}
\end{align}
where we have
\begin{align}
\frac{\partial}{\partial \theta_{j,l}} (\vec{n}_{j,l})_{a_j} = -(\cos(\theta_{j,l})u_{Y,a_j,j,l} + \sin(\theta_{j,l})u_{Z,a_j,j,l}).
\end{align}
The gradient of $f(\vec{\theta}_1, ..., \vec{\theta}_N)$, as defined in the main text, can be computed explicitly and yields

\begin{align}
\frac{\partial}{\partial \theta_{j,l}}f(\vec{\theta}_1, ..., \vec{\theta}_N) = g\sum_{p=1}^L \frac{\partial}{\partial \theta_{j,l}}\operatorname{Tr}(\hat{\rho} h_p)  
-\mu\operatorname{Tr}\left((\hat{M}^{-1})\frac{\partial}{\partial \theta_{j,l}} \hat{M}\right)
\end{align}
where $\frac{\partial}{\partial \theta_{j,l}} \hat{M}$ is computed by taking the element-wise derivative of the matrix elements.

\section{Computational cost}
In this section we discuss the computational cost, measured here as the number of floating point operations necessary for the calculation of the gradient of the cost function. We start with the positivity constraint whose cost function is given by
$ -\mu(t) \operatorname{Tr}\left(\log \left[ \hat{M}(\vec{\theta}_1, ..., \vec{\theta}_N)\right]\right)$. It is straightforward to show that 
\begin{align}
-\mu(t)\frac{\partial}{\partial \theta_{j,l} } \operatorname{Tr}\left(\log \left[ \hat{M}(\vec{\theta}_1, ..., \vec{\theta}_N)\right]\right) =
-\mu(t)\operatorname{Tr}\left(\hat{M}^{-1}\frac{\partial}{\partial \theta_{j,l}} \hat{M}\right).
\end{align}

The matrix $\hat{M}$ is fully parametrized by the set of all Pauli operators of weight $k \leq 2 k_M, r < r_M $ where $k_{M} = 2$ is the maximum Pauli weight and $r_M$ is the maximum range. Each Pauli expectation value can be computed by performing at most $\mathcal{O}(N)$ floating point operations (as they are the average over snapshots of a product of at most $k_M$ floating point numbers). The dimension of the matrix $\hat{M}$ is $\mathcal{O}(L)\times \mathcal{O}(L)$, and it is generally dense. Obtaining all the matrix elements of $\hat{M}$ then requires $\mathcal{O}(N L^2)$ operations. The computational cost to invert the matrix $\hat{M}$ scales as $\mathcal{O}(L^3)$ and is expected to dominate the large $L$ cost. 
Calculating $\frac{\partial}{\partial \theta_{j,l}} \hat{M}$ requires $\mathcal{O}(L)$ floating point operations since only the terms with a Pauli operator at site $j$ which is not the identity contribute (for finite range and finite weight Pauli operators $P_{\vec{a}}$, $P_{\vec{b}}$, the total number of products $P_{\vec{b}}P_{\vec{a}}$ with a non-identity Pauli operator at site $j$ scales as $O(L)$).

Calculating $\operatorname{Tr}\left(\hat{M}^{-1} \frac{\partial}{\partial \theta_{j,l}} \hat{M}\right)$ is equivalent to performing element-wise multiplication between $(\hat{M}^{-1})^T$ and $\frac{\partial}{\partial \theta_{j,l}} \hat{M}$, and then summing the values of the resulting matrix. Since the matrix $\frac{\partial}{\partial \theta_{j,l}} \hat{M}$ contains $\mathcal{O}(L)$ matrix elements which are non vanishing, the number of operations required for this calculation is $\mathcal{O}(L)$. Thus, once $\hat{M}^{-1}$ has been calculated, the cost of computing the derivative with respect to each $\theta_{j,l}$ is $\mathcal{O} \left(L\right)$. 
This calculation must be performed for every site $j$ and every snapshot $l$, bringing the total computational cost to $\mathcal{O}(N L^2)$. 
In total, the number of required floating point operations scales as $\mathcal{O} (L^3)+ \mathcal{O} (N L^2)$. All operations, except for the inversion, can be trivially parallelized between the sites $j$ and the snapshots $l$ to reduce computation time. Incorporating translational invariance can reduce the complexity down to $\mathcal{O} (L)+ \mathcal{O} (NL)$---one can construct all nonequivalent Pauli operators on a unit cell of size at most $r_M$, whose number is at most $4^{r_M}$.  $\mathcal{O} (L)$ correlation matrices of dimension  at most $d \times d$ with $d \sim 4^{r_M}$ can be constructed, which are then individually inverted. As for the gradient calculation, one then has to perform an element-wise multiplication of each inverted block with the corresponding block of the same size, overall incurring an $O(L)$ cost. The gradient only has to be taken over the sites of the unit cell and for each snapshot resulting in an overall $O(NL)$ cost for the gradient.\\

Next, we consider the Hamiltonian cost whose gradient is given by 
\begin{align}
\frac{\partial}{\partial \theta_{j,l}}\hat{H}(\vec{\theta}_1, ..., \vec{\theta}_N) = \sum_{p=1}^L \frac{\partial}{\partial \theta_{j,l}}\operatorname{Tr}(\hat{\rho} h_p).
\end{align}
Assuming that the Hamiltonian terms $h_p$ are composed of Pauli operators whose weight is smaller than or equal to a fixed weight $k_H$ and finite range $r_H$ for any $L$, then the number of floating-point operations necessary to obtain the derivative with respect to a given $\theta_{j,l}$ scales at most as
$\mathcal{O}(1)$ (the derivative selects a single snapshot from the average, and only a finite number of terms overlap non-trivially with the site $j$). This must be computed for every pair $j,l$ bringing the total cost to $\mathcal{O}(NL)$. Here again, this calculation can be parallelized within snapshots and sites, and can be further reduced to $\mathcal{O}(N)$ if translational invariance is employed.

\section{Algorithm}
\subsection{Limited-memory BFGS optimization procedure}
For explicit values of the hyperparameters mentioned in this section, please refer to Tab. \ref{tab:optimization_parameters}.
To accelerate convergence of the variational optimization procedure, we implemented a limited-memory Broyden-Fletcher-Goldfarb-Shanno (L-BFGS) quasi-Newton optimizer \cite{LiuNocedal1989,NocedalWright} combined with a feasibility-preserving backtracking line search \cite{Armijo1966}. The method constructs an implicit approximation to the inverse Hessian using information from a finite number of previous accepted optimization steps, while avoiding the explicit storage or inversion of large matrices. At optimization step $k$, let $\boldsymbol{\theta}_k$ denote the vector of variational parameters and let
\[
\mathbf{g}_k = \nabla_{\boldsymbol{\theta}} f(\boldsymbol{\theta}_k)
\]
be the gradient of the total cost function. After an accepted update step, we define the displacement and gradient-difference vectors
\[
\mathbf{s}_k = \boldsymbol{\theta}_{k+1} - \boldsymbol{\theta}_k,
\qquad
\mathbf{y}_k = \mathbf{g}_{k+1} - \mathbf{g}_k.
\]
Only the most recent $m$ pairs $(\mathbf{s}_k,\mathbf{y}_k)$ are retained in memory, yielding the standard limited-memory formulation \cite{LiuNocedal1989}. To ensure numerical stability and preserve positive-definiteness of the approximate inverse Hessian, curvature updates are accepted only if
\[
\mathbf{s}_k^\top \mathbf{y}_k
>
\epsilon_{\mathrm{curv}}
\,
\|\mathbf{s}_k\|
\,
\|\mathbf{y}_k\|,
\]
where $\epsilon_{\mathrm{curv}}$ is a small positive threshold. Updates failing this condition are discarded and are not added to the L-BFGS history. The search direction is computed using the standard two-loop recursion \cite{LiuNocedal1989,NocedalWright}. Starting from
\[
\mathbf{q} = \mathbf{g}_k,
\]
we first apply the stored curvature corrections in reverse chronological order:
\[
\alpha_i
=
\rho_i \mathbf{s}_i^\top \mathbf{q},
\qquad
\mathbf{q}
\leftarrow
\mathbf{q} - \alpha_i \mathbf{y}_i,
\]
with
\[
\rho_i = \frac{1}{\mathbf{y}_i^\top \mathbf{s}_i}.
\]
The resulting vector is then scaled by an initial inverse-Hessian approximation,
\[
\mathbf{r} = \gamma_k \mathbf{q},
\qquad
\gamma_k
=
\frac{\mathbf{s}_{k-1}^\top \mathbf{y}_{k-1}}
{\mathbf{y}_{k-1}^\top \mathbf{y}_{k-1}},
\]
where $\gamma_k$ provides a scalar estimate of the local inverse curvature. A second recursive pass is performed in chronological order:
\[
\beta_i
=
\rho_i \mathbf{y}_i^\top \mathbf{r},
\qquad
\mathbf{r}
\leftarrow
\mathbf{r}
+
\mathbf{s}_i(\alpha_i-\beta_i).
\]
The proposed quasi-Newton direction is then
\[
\mathbf{p}_k = -\mathbf{r}.
\]
As a safeguard, we explicitly verify that this is a descent direction,
\[
\mathbf{g}_k^\top \mathbf{p}_k < 0.
\]
If this condition is violated, the algorithm discards the L-BFGS direction and falls back to steepest descent,
\[
\mathbf{p}_k = -\mathbf{g}_k.
\]
The final step size is chosen using a backtracking line search. Starting from an initial step size $\eta_0$, we consider trial updates
\[
\boldsymbol{\theta}_{k+1}^{\mathrm{trial}}
=
\boldsymbol{\theta}_k
+
\eta \mathbf{p}_k.
\]
A trial point is accepted only if it satisfies the imposed feasibility constraints and obeys the Armijo sufficient-decrease condition \cite{Armijo1966}
\[
f(\boldsymbol{\theta}_k+\eta\mathbf{p}_k)
\le
f(\boldsymbol{\theta}_k)
+
c\,\eta\,\mathbf{g}_k^\top \mathbf{p}_k,
\]
where $c$ is the Armijo parameter. If either feasibility or sufficient decrease fails, the step size is reduced according to
\[
\eta \leftarrow \tau \eta,
\]
with $0<\tau<1$, and the trial is repeated until either a step is accepted or a minimum allowed step size is reached. If no feasible decreasing step is found, the parameters are restored to the last accepted point and the L-BFGS memory is reset. The memory is also reset when the accepted step size becomes extremely small, since in this regime the stored curvature information may no longer provide a reliable approximation to the local geometry. Below is a table with the values of all the constants mentioned above:

\begin{table}[t]
\centering
\caption{Optimization and L-BFGS hyperparameters used throughout the simulations.}
\begin{tabular}{lll}
\hline
\textbf{Parameter} & \textbf{Symbol / Name} & \textbf{Value} \\
\hline

Maximum Pauli weight & $w_M$ & $2$ \\

Maximum Pauli range & $r_M$ & $2 \text{ to }4$ \\

Total optimization epochs & $T$ & $300$ \\

Initial learning rate & $\eta_0$ & $1$ \\

Reset learning rate threshold & $\eta_{\text{threshold}}$ & $10^{-8}$ \\

Backtracking decay factor & $\tau$ & $0.5$ \\

L-BFGS memory depth & $m$ & $8$ \\

Curvature threshold & $\epsilon_{\mathrm{curv}}$ & $10^{-5}$ \\

Armijo parameter & $c$ & $10^{-4}$ \\

Hamiltonian scaling schedule & $g(t)$ & $1/L$ \\

Final PSD penalty scale & $A_{\mathrm{final}}$ & $5\times10^{-4}$ \\

PSD scaling schedule &
$\mu(t)$ &
\(
\displaystyle
\frac{1}{\operatorname{Tr}(M)}
\left(A_{\mathrm{final}}\right)^{t/T}
\)

\\

\hline
\end{tabular}
\label{tab:optimization_parameters}
\end{table}

\subsubsection{Optimization procedure}

The optimization of the angles $\theta_{j,l}$ proceeds as follows. For a given Hamiltonian $H$, a system size $L$ and a number of snapshots $N$, the angles $\theta_{j,l}$ are set uniformly in the range $[0,2\pi]$, and a Haar random unitary $U_{j,l}$ is sampled uniformly for each pair $j,l$. Let us recall that the cost function is given by 

\begin{align}
    f(\vec{\theta}_l, t) = g \hat{H} (\vec{\theta}_1, ..., \vec{\theta}_N) - \mu(t) \text{Tr} \log \left[ \hat{M} (\vec{\theta}_1, ..., \vec{\theta}_N)\right]. 
\end{align}

The initial randomly selected values of $\vec{\theta}$ usually generate a corresponding matrix $\hat{M} (\vec{\theta}_1, ..., \vec{\theta}_N)$ whose eigenvalues are negative. As such, a warm-up optimization is required to push the eigenvalues above $0$ in order to use interior-point methods. To achieve this, we start by optimizing the modified cost function 

\begin{align}
    f_{\text{pre}}(\vec{\theta}_l, t) = \frac{-1}{\operatorname{Tr}(M)} \text{Tr} \log \left[ \hat{M} (\vec{\theta}_1, ..., \vec{\theta}_N) + \Delta(t)\right]. 
\end{align}

Note that $\operatorname{Tr}(M) = \text{dim}(M)$ and is thus independent of the angles. At each time step $t$ the function $\Delta(t)$ is set to $\Delta(t) = -\lambda_{\text{min}}(t) + \delta$ where $\delta = 0.01$ and $\lambda_{\text{min}}(t)$ is the minimum eigenvalue of $M$ at epoch $t$. Importantly, we treat $\Delta(t)$ as a constant introduced at each epoch $t$, and thus its implicit dependence on $M$ is not included in the gradient calculation. This approach generates a consistent upward push on the eigenvalues of $M$, which we find typically produces a fully positive semi-definite matrix in about $10$ to $20$ epochs for most models. The number of required steps can be larger if the number of snapshots is small and can vary based on the number of imposed PSD constraints. Once the preoptimization is completed we set $g(t) = \frac{1}{L}$ and $\mu(t) = \frac{1}{\text{Tr}(M)}A_{\text{final}}^{t/T}$ where $t \in [0,1,...,T]$, $T = 300$ and $A_{\text{final}} = 5\times 10^{-4}$ and perform the L-BFGS optimization discussed in the previous section.

\end{widetext}
\bibliographystyle{apsrev4-1}
\bibliography{HLearning_arxiv_V2}
\end{document}